\documentclass{emulateapj}

\usepackage{macros}

\begin{document}

\title{A Multiwavelength View of a Mass Outflow from the Galactic Center}
\shorttitle{Mass Outflow from GC}
\shortauthors{Law}

\author{C. J. Law}
\affil{Radio Astronomy Lab, University of California, Berkeley, CA 94720; claw@astro.berkeley.edu}

\begin{abstract}
The Galactic center (GC) lobe is a degree-tall shell of gas that spans the central degree of our Galaxy.  It has been cited as evidence for a mass outflow from our GC region, which has inspired diverse models for its origin.  However, most work has focused on the morphology of the GC lobe, which has made it difficult to draw strong conclusions about its nature.  Here, I present a coherent, multiwavelength analysis of new and archival observations of the GC lobe.  New radio continuum observations show that the entire structure has a similar spectral index, indicating that it has a common origin.  The radio continuum emission shows that the GC lobe has a magnetized layer with a diameter of 110 pc and an equipartition field strength ranging from 40 to 100 $\mu$G.  I show that optical and radio recombination line emission are associated with the GC lobe and are consistent with being located in the GC region.  The recombination line emission traces an ionized shell nested within the radio continuum with diameter of 80 pc and height 165 pc.  Mid-infrared maps at 8 and 15 $\mu$m show that the GC lobe has a third layer of warm dust and PAH-emission that surrounds the radio continuum shell with a diameter of 130 pc.  Assuming adibatic expansion of the gas in the GC lobe, its formation required an energy input of about $5\times10^{52}$ ergs.  I compare the physical conditions of the GC lobe to several models and find best agreement with the canonical starburst outflow model.  The formation of the GC lobe is consistent with the currently observed pressure and star formation rate in the central tens of parsecs of our Galaxy.  Outflows of this scale are more typical of dwarf galaxies and would not be easily detected in nearby spiral galaxies.  Thus, the existence of such an outflow in our own Galaxy may indicate that it is relatively common phenomenon in the nuclei of spiral galaxies.
\end{abstract}


\section{Introduction}
The GC lobe is a shell of emission rising north of the Galactic plane to a height of roughly 1\sdeg \citep[$\sim$140 pc at 8 kpc][]{r93}.  As shown in Figure \ref{gcl620}, it spans the central degree of the GC region, a region noteworthy for a massive black hole, star clusters, and exotic objects \citep{g05,f99,y88}.  It was discovered in a 10 GHz radio continuum survey of the GC region \citep{s84}.  The discovery observations measured a thermal spectral index at 10 GHz, implying that it had an ionized mass of $4\times10^5$ \msol\ \citep{s85}.  The shape and location of the shell motivated several models for its formation, including starburst \citep{c92}, AGN \citep{m01}, and magnetodynamic effects \citep{u85}.  

Each model for the formation of the GC lobe has important implications for our understanding of galactic nuclei.  If it is formed by a starburst, it tells us about the star formation history of the GC region.  Is such activity episodic \citep{st04} or persistent \citep{f04}?  If the GC lobe was formed by a jet, it would tell us that Sgr A* can sometimes have powerful outflows, as opposed to its relatively quiescent current state \citep{mar05}.  Knowledge of such activity would have implications for AGN feedback in galaxy evolution \citep{h05}.  The presence of magnetodynamic effects in the GC region would alter our interpretation of the energetics of outflows in other galaxies \citep{su87}.

However, later observations complicated the interpretation that the GC lobe was an outflow in the GC region.  Radio continuum observations measured a nonthermal spectral index in the east \citep{r87} and thermal emission in the west \citep{b03b}.  This was consistent with differences in the polarization fraction of the east and west sides of the lobe, which found the east side much more strongly polarized \citep{t86,h92}.  Furthermore, there was an unusual shocked molecular cloud (named AFGL 5376) associated only with the western half of the GC lobe, implying the two halves had different origins \citep{u90}.  These asymmetries between east and west sides of the GC lobe cast doubts on the idea that it was a coherent structure.

The idea that the GC lobe was a coherent object was reinforced with the detection of a remarkable mid-IR shell \citep{b03}.  The \emph{Midcourse Space Experiment} \citep[\emph{MSX};][]{p01} survey of the Galactic center at 8 $\mu$m found filamentary structures coincident with the entire GC lobe.  \citet{b03} modeled the mid-IR emission as shock-heated dust entrained in an outflow.  This model implies that the outflow has a mass of $5\times10^5$ \msol\ and (assuming an expansion velocity) a kinetic energy greater than $10^{54}$ ergs.  This mass and energy are comparable to those observed in small- to moderate-sized starburst outflows \citep{v05}.

Despite the strong morphological evidence in radio continuum and mid-IR observations, there are several lingering questions about the nature of the GC lobe.  What created the shell?  Could it be in the foreground to the GC region?  Can it be related to other GC objects?  To answer these questions in detail, I analyzed new and archival observations of the GC lobe, including new radio continuum, optical recombination line, and mid-IR observations.  Section \ref{obs} describes the results of each observation.  Section \ref{points} summarizes a few simple conclusions about the nature of the GC lobe.  The morphology and physical conditions in the lobe show that it is a layered, three-dimensional shell of gas in the GC region.  Section \ref{discussion} discusses models for the formation of the GC lobe and how it relates to other GC objects.  I find that the GC lobe is best described as an outflow powered by current star formation.  Section \ref{conclusions} presents the conclusions of this study.

\section{Observations and Results}
\label{obs}

\subsection{Radio Continuum}
\label{all_gbtcont}
\citet{gcsurvey_gbt} present a survey of the central two degrees of the Galactic plane with the GBT at 3.5, 6, 20, and 90 cm.  Here, I discuss the first results from this survey related to the GC lobe, including morphological, spectral, and other properties of the radio continuum emission.
 
\begin{figure}[tbp]
\includegraphics[width=\textwidth]{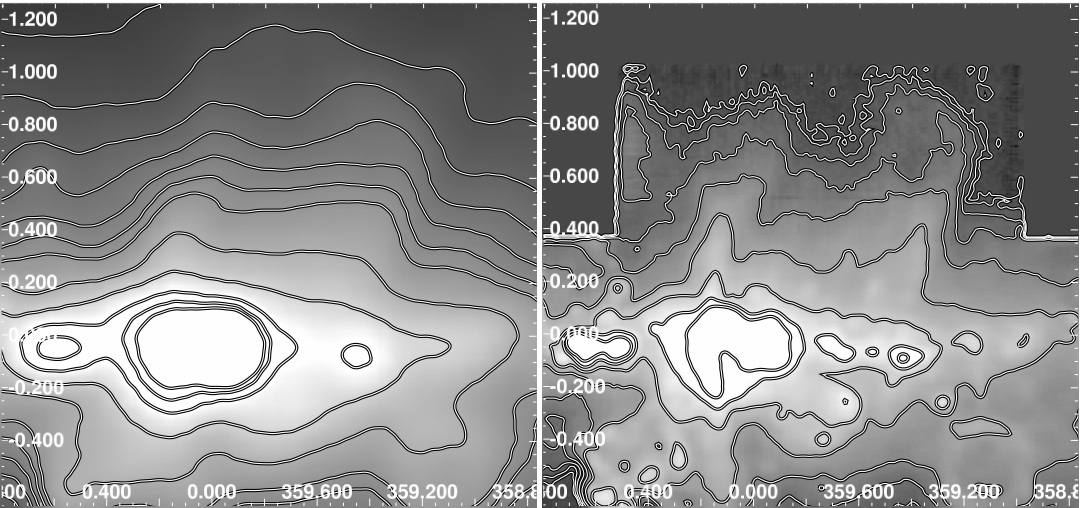}
\caption{\emph{Left}: 20 cm GBT image of the central degree of the Galaxy.  The GC lobe appears as two columns of emission at $l=359\ddeg3$ and $0\ddeg1$; at 20 cm, the columns are most visible north of $b=0\ddeg3$.  The 20 cm map also shows a weak cap to the GC lobe near $b=1$\sdeg.  Contours are at levels of 12, 14, \ldots, 24, 30, 40, 60, 80, 100 Jy beam$^{-1}$.  \emph{Right}: 6 cm GBT image of the same region;  the top of the GC lobe is not covered by this survey.  Contours are at levels of $0.02*2^n$ Jy beam$^{-1}$, with $n=0-8$.  \label{gcl620}}
\end{figure}

The 6 and 20 cm surveys covered the brightest part of the GC lobe and are shown in Figure \ref{gcl620}.  The observed morphology is brightest at the eastern and western edges, but also shows a cap at the northern edge in the 20 cm image.  As discussed in \citet{b03}, the radio continuum shell is very tightly correlated with a shell of mid-IR emission.  The 6 cm image in Figure \ref{gcl620} shows how the emission at the edges of the GC lobe are much broader north of $b=0\ddeg5$, forming a forked or fan-like structure of width $\sim0\ddeg3$.  The distance from the east to the west side of the GC lobe is $0\ddeg8$, equivalent to 110 pc at the GC.  The height of the GC lobe from the Galactic plane is roughly $1\ddeg2$, or 165 pc at the GC.

\begin{figure}[tbp]
\includegraphics[width=\textwidth]{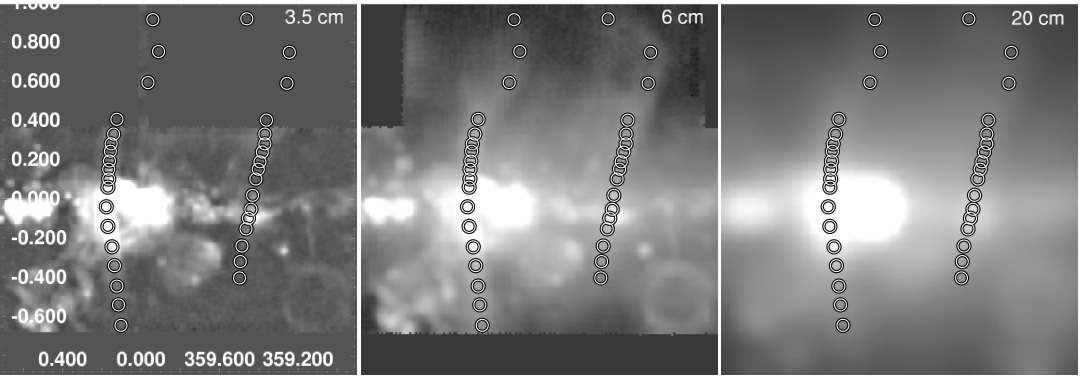}
\caption{GBT images showing the GC lobe at 3.5, 6, and 20 cm.  Circles in each image show the approximate location at which spectral index measurements were made.  Note that at some locations, the GC lobe is confused with other well-known GC objects, such as Sgr C or the Arches;  these areas are labeled in Fig. \ref{gclspix}. \label{gclspixpos}}
\end{figure}

The spectral index was measured by taking slices across the 3.5, 6, and 20 cm images convolved to the same resolution.  \citet{gcsurvey_gbt} describes the spectral index technique and its application to other GC objects.  Figure \ref{gclspixpos} shows the locations where the spectral index was measured in the GC lob, corresponding to the peak flux along the eastern and western edges.

\begin{figure}[tbp]
\begin{center}
\includegraphics[width=\textwidth]{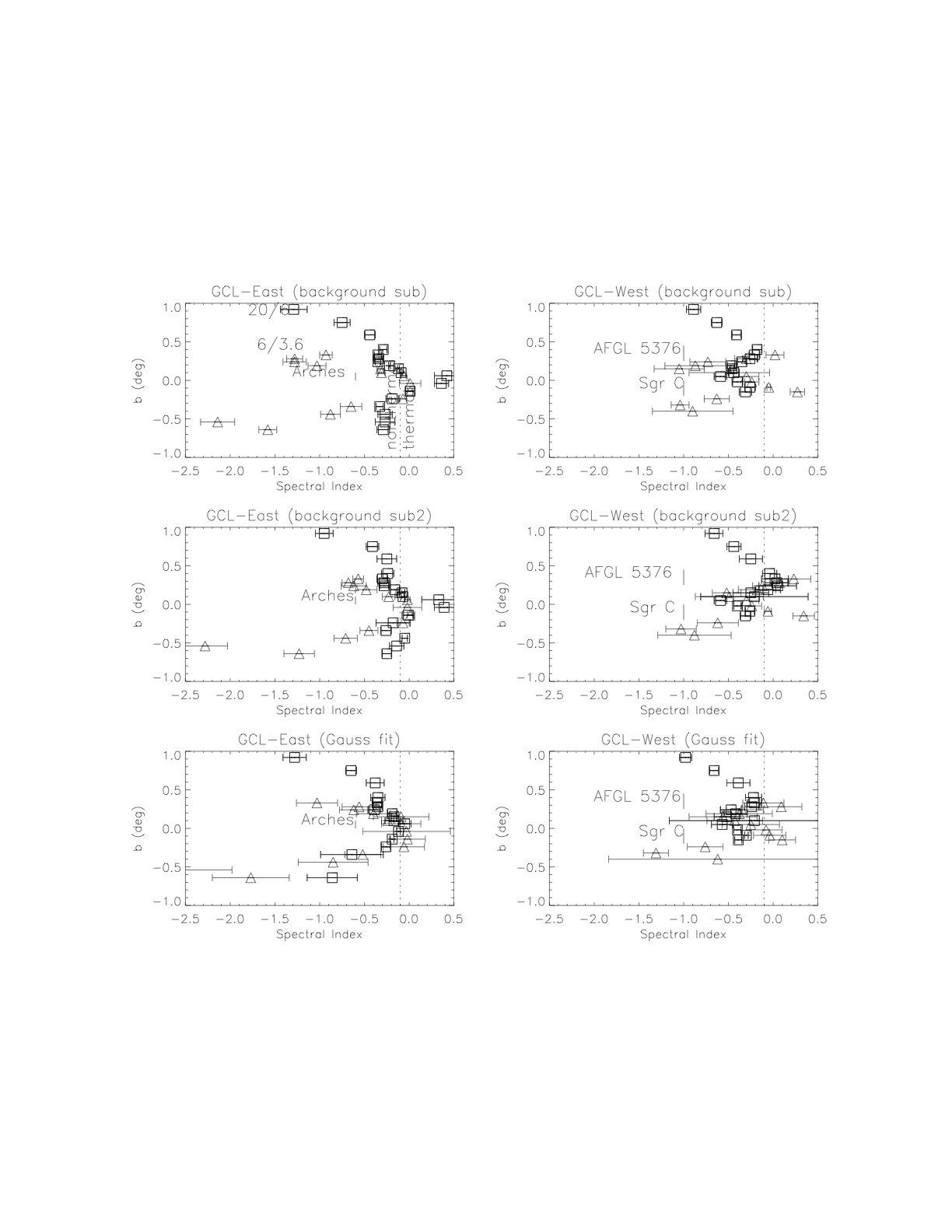}
\end{center}
\vspace{-5cm}
\caption{\emph{Top panels}:  The 20/6 cm and 6/3.5 cm spectral indices (``$\alpha_{20/6 cm}$'' and ``$\alpha_{6/3.5 cm}$'', respectively) of the GC lobe as a function of Galactic latitude for the eastern and western halves of the GC lobe.  These panels show the results of a background-subtraction method for estmiating the peak brightness.  Square symbols show the 20/6 spectral index and triangles show the 6/3.5 spectral index.  The vertical dotted line shows a thermal-like spectral index of --0.1;  a nonthermal spectral index lies left of this line.  Ranges in galactic latitude where the GC lobe might be confused with other sources are labeled with vertical bars and the name of the confusing source.  \emph{Middle panels}:  Same as for the top panels, but with a background estimated close to the peak brightness of the GC lobe. \emph{Bottom panels}:  Same as for the top panels, but using the Gaussian-fit technique. \label{gclspix}}
\end{figure}

The spectral indices measured for the GC lobe at these locations are shown in Figure \ref{gclspix}.  To get a robust measurement, three different assumptions are used in these measurements.  The top two rows show the spectral index measurement when the background was fit with a first order polynomial.  The spectral index was calculated assuming a background very close to the peak emission (shown in the top row) and at the edges of the slice (shown in the middle row).  The bottom row shows a measurement of the spectral index when a model of a linear background plus a Gaussian is fit to the slice.  All structure in the slice that does not match this model and does not appear to be noise (e.g., unrelated sources) is ignored.  The uncertainty in the brightness is estimated by calculating the rms deviation of the fit residuals.

Using a range of techniques to measure the spectral index in the GC lobe gives confidence in the trends that are common between them.  Inspecting Figure \ref{gclspix} reveals several trends in the spectral index of the GC lobe.

\begin{enumerate}
 \item The values of $\alpha_{6/3.5 cm}$ and $\alpha_{20/6 cm}$ are significantly nonthermal, where not confused with other sources.  In the east, all spectral indices north of the Arches \citep{l01} and south of $b\sim-0.3$ are significantly less than an optically thin thermal index of $-0.1$.  In the west, the indices are consistent with a nonthermal origin, except at Sgr C and AFGL 5376.  \footnote{These values are different from \citet{s85}, since that work measured the spectral index from images that had been unsharp masked \citep[``BGF technique'';][]{r87}.  Since the central degrees of the Galaxy are filled with nonthermal emission, an unsharp mask tends to subtract more flux at long wavelengths.  This effect likely explains why \citet{s85} found flatter spectral index than that shown here or in other works \citep{p92}.}
 \item There is a clear decrease of $\alpha_{20/6 cm}$ at large positive latitudes in both the east and west.  All three methods show that the indices in the northernmost three slices are nonthermal and progressively more negative toward higher latitudes.  Previous observations showed this for the eastern half of the GC lobe \citep{p92}.
 \item There is a decrease in $\alpha_{6/3.5 cm}$ at large negative latitudes in the east and west.  In the east, the decrease is significant, if irregular.  In the west, the decrease is less clear, due to confusion with Sgr C, but it appears to decrease in all three analysis techniques.
 \item $\alpha_{20/6 cm}$ tends to be flatter than $\alpha_{6/3.5 cm}$ far from plane in the east.  North of the Arches and south of $b\sim-0.3$, $\alpha_{6/3.5 cm}$ seems to decrease continually.  In contrast, $\alpha_{20/6 cm}$ seems to have a fairly regular value of $\sim-0.3$ (particularly noticeable in the top panels of Figure \ref{gclspix}).
 \item The values of $\alpha_{20/6 cm}$ and $\alpha_{6/3.5 cm}$ are flatter near AFGL 5376.  The $\alpha_{6/3.5 cm}$ index is consistent with thermal emission, but $\alpha_{20/6 cm}$ is significantly nonthermal.  The middle panels show that $\alpha_{20/6 cm}$ is thermal, but this technique generally biases the index upward in confused regions.
\end{enumerate}

\begin{figure}[tbp]
\includegraphics[width=\textwidth]{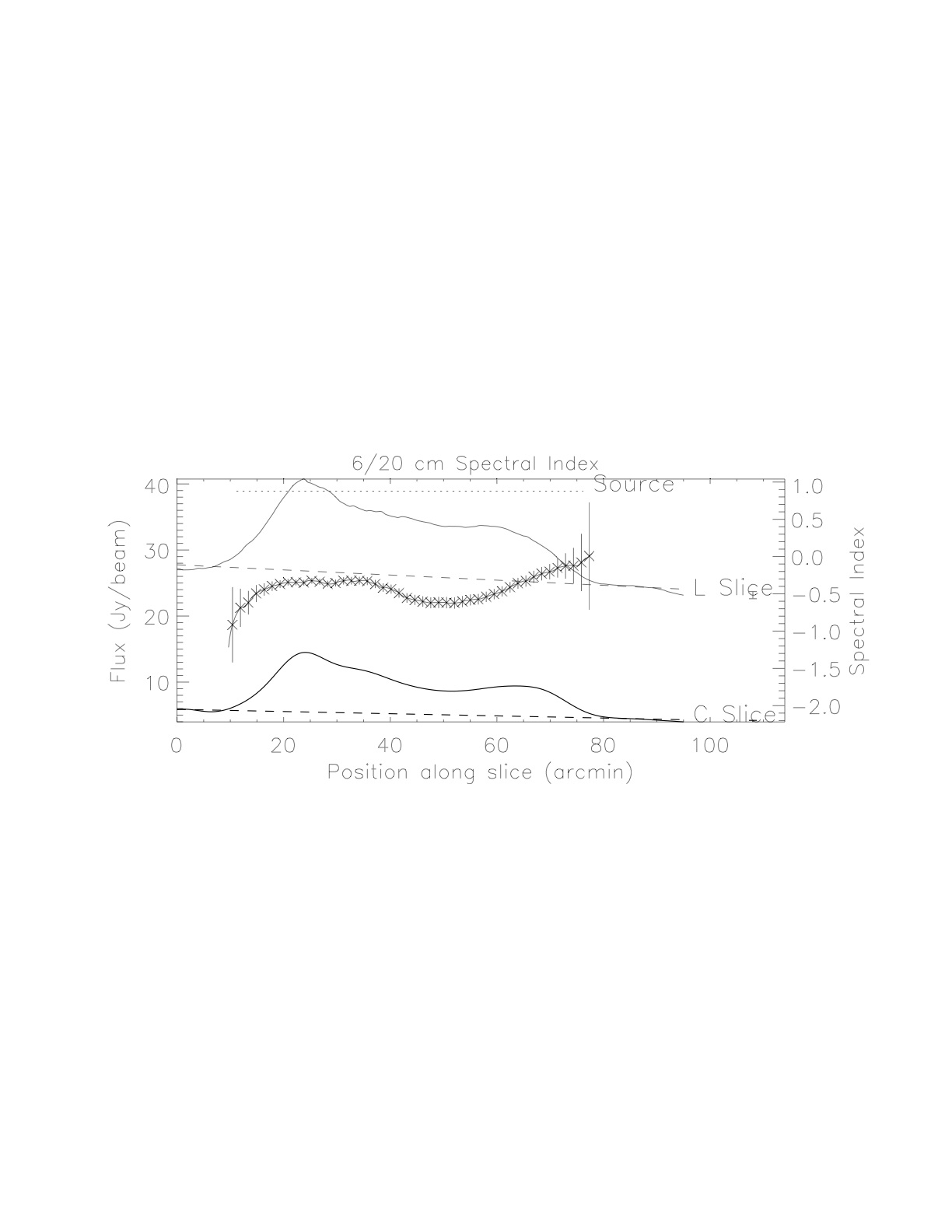}
\vspace{-3cm}
\caption{Plot of the brightness at 6 cm (``C band'', bottom) and 20 cm (``L band'', top) for a slice across the GC lobe at $b=0\ddeg24$, convolved to the same resolution.  The spectral index for the GC lobe emission measured from the slices is shown with crosses.  The dashed line shows the best-fit background line fit to each slice, ignoring the GC lobe emission (shown with the dotted line).  To reduce confusion, only points with spectral index error less than 1 dex are plotted.  The spectral index is highest near the peaks of the 6 and 20 cm brightness and lowest between the peaks (i.e., ``inside'', assuming a shell geometry). \label{gclspixslice}}
\end{figure}

Figure \ref{gclspixslice} shows the flux and spectral index of a slice across the GC lobe at $b=0\ddeg24$ using the background-fit technique.  This slice shows that the spectral index increases near the peak brightness; slices at all latitudes north of the Galactic plane show the same behavior.   Figure \ref{gclspixslice} also shows that the spectral index tends to increase toward the ``outside'' of the GC lobe (assuming a shell geometry).  Moving eastward across the east edge and westward across the west edge shows a slight increase in the spectral index.  Another way of saying this is that the peak of the 6 cm emission tends to lie outside of the 20 cm emission.  

From these trends, I conclude that the radio continuum emission associated with the GC lobe is dominated by synchrotron emission.  This confirms the work of \citet{p92} and extends it to include the entire radio continuum structure of the GC lobe.  I estimate the magnetic field using the revised equations of \citet{b05}, assuming a shell-like geometry with a path length through the edge of 50 pc (see \S\ \ref{shellsec}), a proton-to-electron energy density ratio relative to the filling factor (i.e., $K/f$) of 100, and integrating over energies from 10 MHz to 10 GHz.  Under these assumptions, the typical GC lobe emission at $b=0\ddeg6-0\ddeg9$ has a field strength of 38--53 $\mu$G.  For the GC lobe emission at $0\ddeg2<|b|<0\ddeg5$, the revised equipartition magnetic field equation is 100 $\mu$G.  These minimum magnetic field strengths for the GC lobe are similar to that estimated elsewhere in the GC region \citep{r87,la05}.

\subsection{Optical H$\alpha$}
\begin{figure}[tbp]
\includegraphics[width=\textwidth]{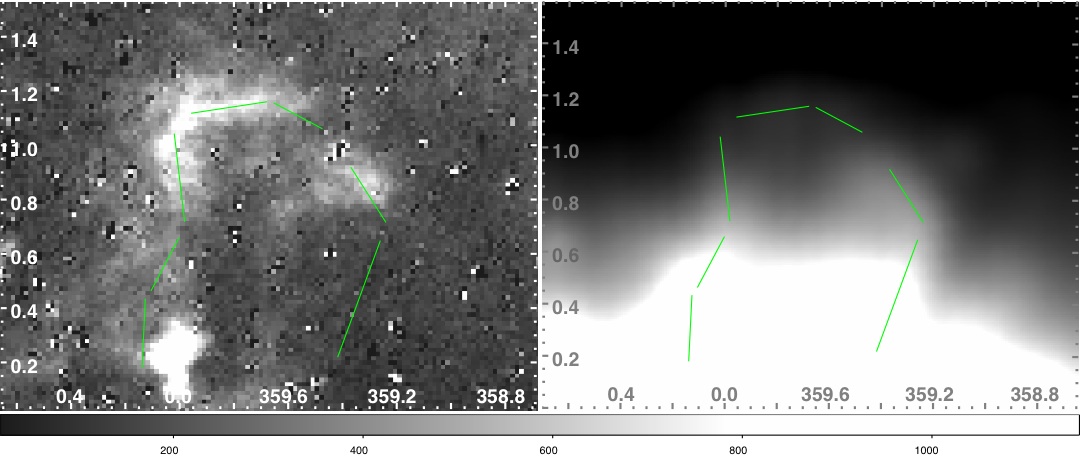}
\caption{\emph{Left}: SHASSA continuum-subtracted H$\alpha$ image toward the GC lobe with Galactic coordinates.  The bar at bottom shows the intensity scale in deciRayleighs. \emph{Right}:  GBT 20 cm radio continuum image, similar to that shown in Fig. \ref{gcl620}.  For reference, the lines in both panels traces the shell of the radio continuum emission observed at 20 cm.  \label{shassa}}
\end{figure}

The GC region was surveyed in the hydrogen Balmer $\alpha$ transition as a part of the Southern H-$\alpha$ Sky Survey Atlas (SHASSA) project \citep{g01}.  Figure \ref{shassa} shows the portion of that survey coincident with the GC lobe.  The arcminute-resolution map shows filamentary emission tracing the northern part of the GC lobe.  The association of optical H-$\alpha$ emission with the GC lobe has not been discussed before.  What does this imply for the origin of the ionized gas?

\citet{gcl_recomb} find radio recombination line emission clearly associated with the GC lobe, which constrains the properties of its ionized gas.  That works shows that the thermal free-free emission has a maximum brightness temperature of about 0.6 K at 4.9 GHz.  The brightness temperature and electron temperature constrain the the H$\alpha$ line flux \citep{g01}, giving an upper limit on the expected H$\alpha$ flux, $I_{\rm{H}\alpha}^{\rm{pred}}\lesssim2000$ Rayleighs ($1\ \rm{R} = 2.41\times10^{-7}$ erg cm$^{-2}$ s$^{-1}$ sr$^{-1}$ for H$\alpha$).  The background-subtracted H$\alpha$ line flux seen by SHASSA is about 4 R.  If optical and radio recombination lines are coming from the same gas, the flux difference implies an optical extinction factor of at most 500, or $\tau_{\rm{H}\alpha}\lesssim6.2$.  For a typical interstellar extinction law \citep[$R_V=3.1$]{c89,mad05}, this corresponds to a visual extinction of $A(V)\lesssim8.7$ mag. \footnote{It is worth noting that the electron temperature upper limit is unusually small and the line likely has little Doppler broadening.  If so, the predicted upper limit to the H$\alpha$ flux and extinction are likely close to the true flux and extinction.}

Is the implied extinction consistent with expectations for this region?  \citet{d03} mapped the extinction in the central several degrees and showed that at the north of the GC lobe, the typical extinction is $A(K)\approx1$ mag or $A(V)\approx9$ mag \citep{c89}.  Considering the uncertainty in the extinction law in this region \citep{g09}, the known extinction is consistent with the extinction required to explain the H$\alpha$ flux.  Thus, the GC lobe's ionized gas seen in radio recombination line emission can also explain the coincident H$\alpha$ emission.  Furthermore, the sensitivity of SHASSA ($\sim2$ R) implies an extinction limit of $A(V)\sim10$ mag for this emission.  This $A(V)$ is typical south of $b=+1$\sdeg, which explains why no H$\alpha$ is detected from the rest of the GC lobe.

\subsection{Mid-IR Continuum}
\begin{figure}[tbp]
\includegraphics[width=\textwidth]{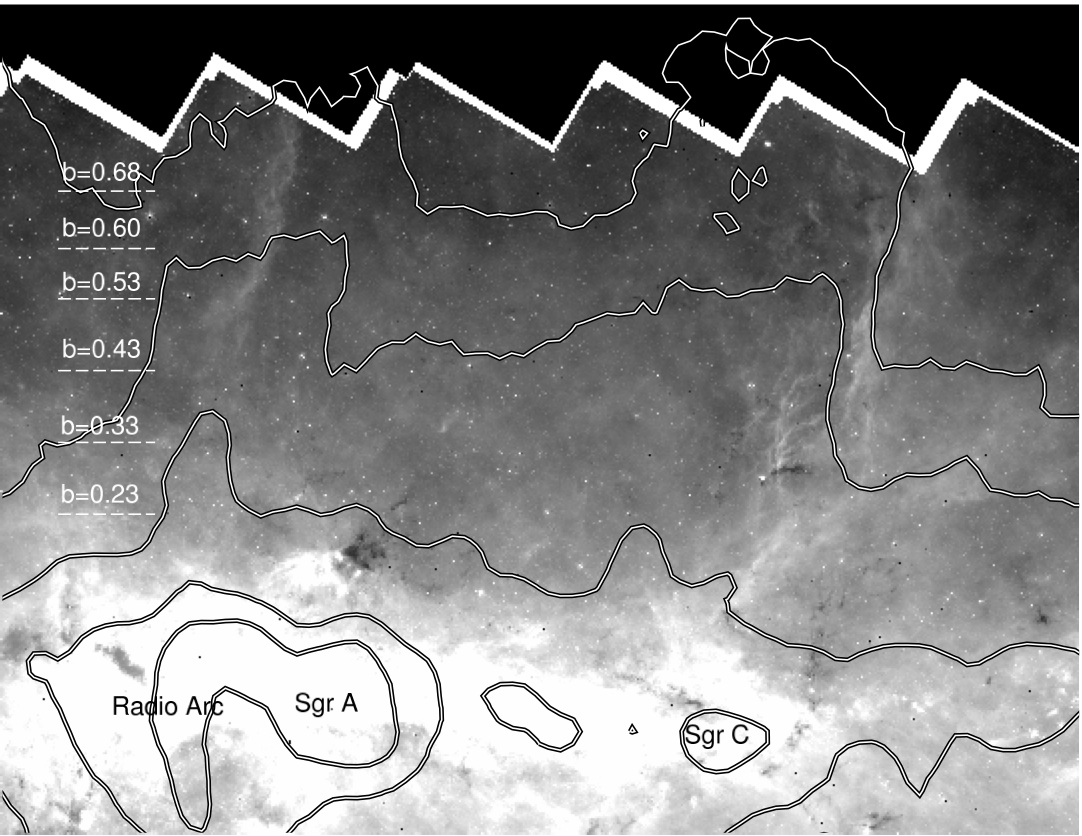}
\caption{Grayscale shows a portion of the \spitzer/IRAC 8 $\mu$m image and contours show GBT 6 cm continuum emission around the GC lobe.  Dashed lines show the latitudes at which slices were taken across the GC lobe for Fig. \ref{slices}; only a portion of the slices are shown here to avoid confusing the image.  \label{sp4GCC}}
\end{figure}

The central two degrees of our Galaxy was surveyed with \spitzer/IRAC in four bands from 2 to 8 $\mu$m \citep{s06,r08,a08}.  Figure \ref{sp4GCC} shows how the \spitzer\ 8 $\mu$m emission correlates with the 6 cm radio continuum emission of the GC lobe.  The correlation has been noted \citep{b03}, but a quantitative comparison of the various components has not yet been done.

\begin{figure}[tbp]
\includegraphics[width=0.5\textwidth]{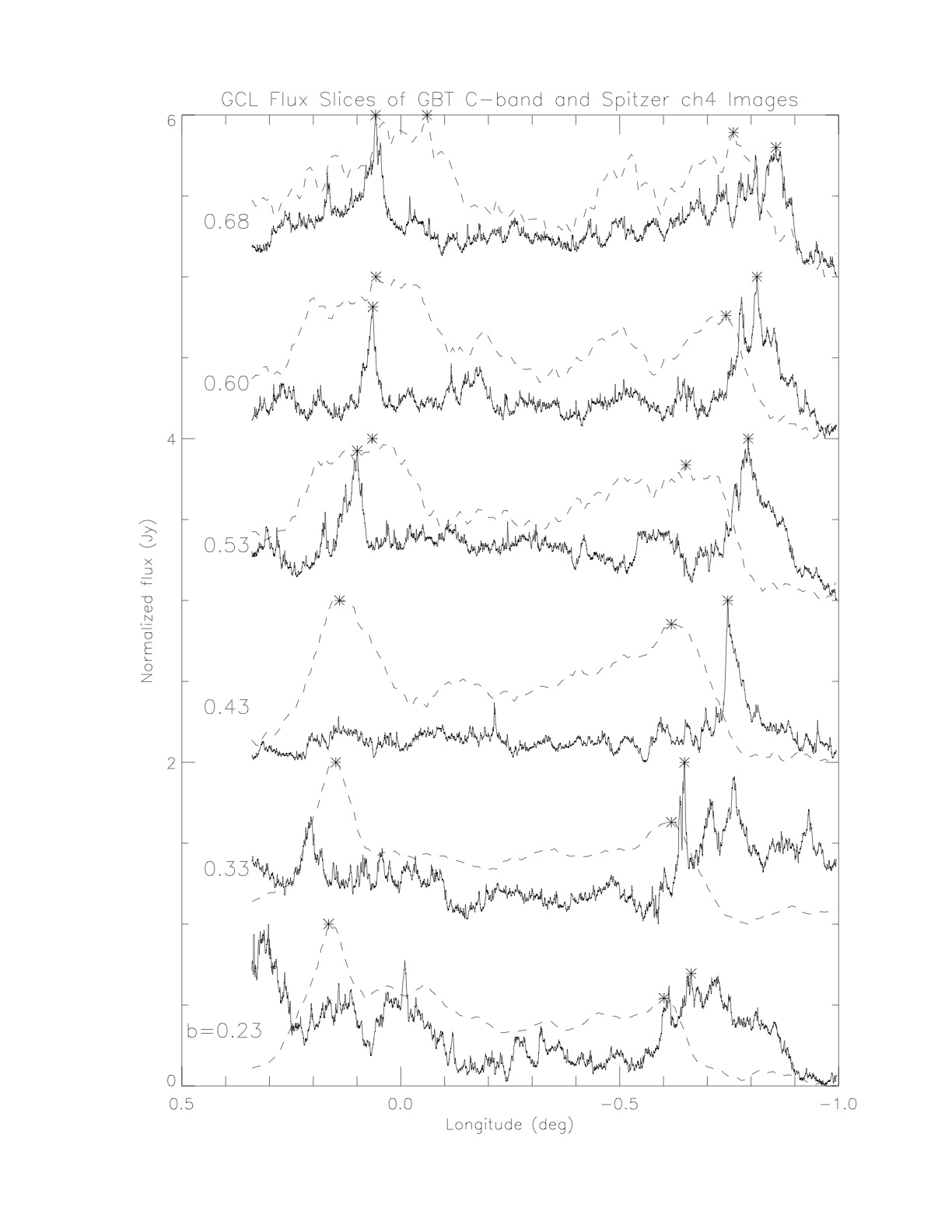}
\includegraphics[width=0.5\textwidth]{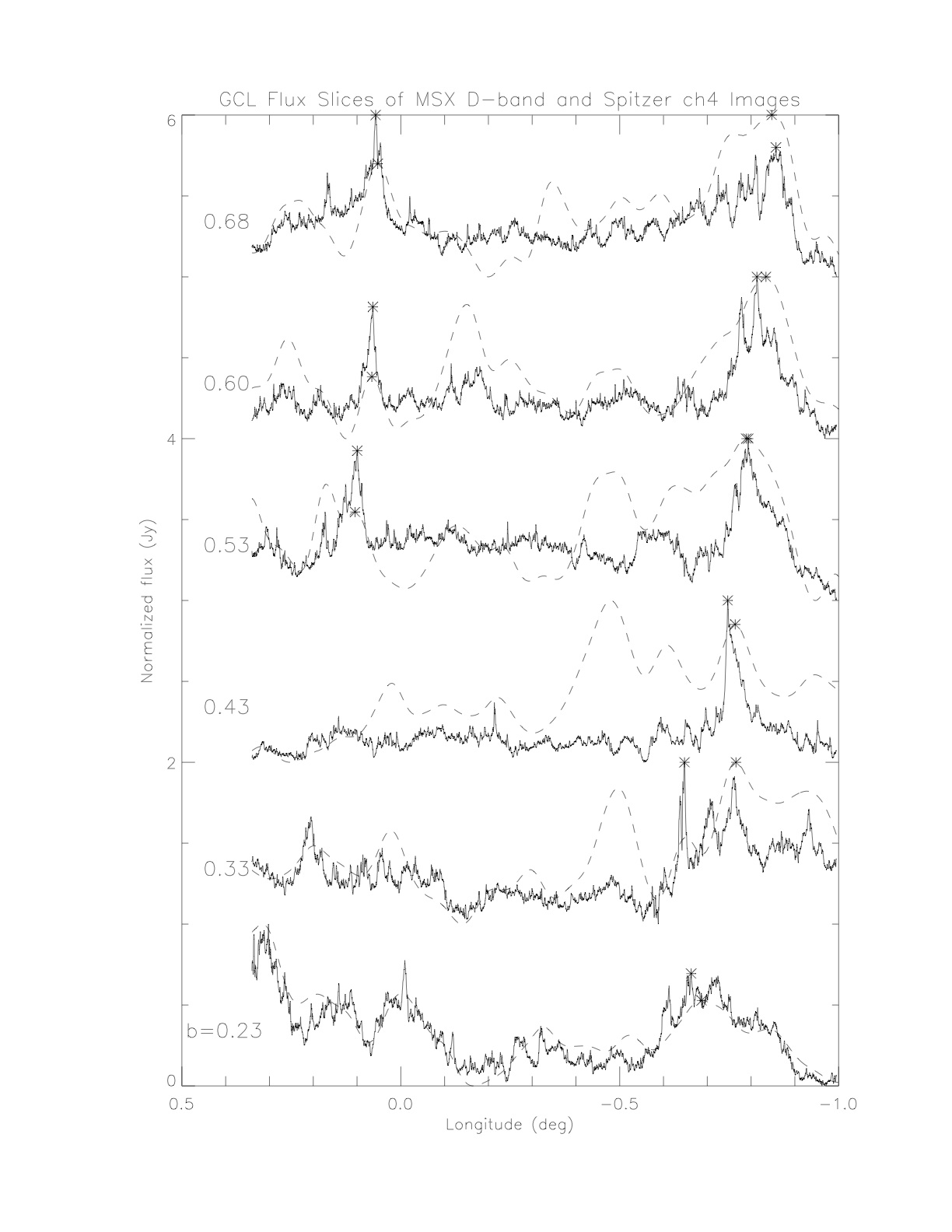}
\caption{\emph{Left}:  Normalized flux slices of GBT 6 cm continuum (dashed line) and \spitzer/IRAC 8 $\mu$m (solid line) for six latitudes across the GC lobe.  The bottom slices show the lowest latitude and subsequent slices are vertically offset by 1 for clarity.  The slices are over identical longitude ranges at a single galactic latitude, which is indicated for each pair.  The slices were calculated for a small range of latitudes and median filtered (by pixel) to remove compact sources.  The peak flux for the eastern and western halves of each slice are shown with stars, where they represent the peak flux of the GC lobe.  The lower latitude slices do not have a clear GC lobe counterpart in the mid-IR on the west side, so no stars are plotted there.  The slices show that the peak of the 20 cm emission is shifted relative to the 8$\mu$m emission at all latitudes.  \emph{Right}: Same as the left side, but comparing smoothed \msx\ 15 $\mu$m (``D-band'') slices to the \spitzer/IRAC 8 $\mu$m slices. The slices show that the peaks of the 8 $\mu$m and 20 $\mu$m emission are roughly coincident.  \label{slices}}
\end{figure}

Figure \ref{slices} compares slices across the GBT radio continuum, \spitzer\ 8 $\mu$m, and \msx\ 15 $\mu$m maps.  The fluxes for slices across six latitudes are normalized and plotted together to show the relative location of the peak flux throughout the lobe.  The \spitzer\ 8 $\mu$m band is dominated by PAH emission features \citep{a08}, so it traces dust column density and the UV radiation field \citep{p04}.  The \msx\ 15 $\mu$m band does not have any strong PAH or line emission, so it is a good tracer of dust continuum emission \citep{a89}.

The slices confirm that the radio continuum and 8 $\mu$m peaks are offset from each other, with the nonthermal radio continuum emission inside the 8 $\mu$m emission.  The positional offset is typically $0\ddeg05$ in the east and $0\ddeg1$ in the west for the range of latitudes where 8 $\mu$m emission is seen ($b=0\ddeg23$--$0\ddeg68$).  The peak-to-peak width of the 8 $\mu$m shell is $0\ddeg9$, which is $\approx130$ pc at the GC distance.

The right side of Figure \ref{slices} compares the \msx\ 15 $\mu$m flux to the \spitzer\ 8 $\mu$m flux.  In general, the peak of the GC lobe structure in \msx\ 15 $\mu$m image is very near the peak of \spitzer\ 8 $\mu$m image.  Since the \msx\ image shows warm dust and \spitzer\ image shows PAH emission, I conclude that the GC lobe is host to warm dust and is being irradiated by far-UV photons.

\section{Conclusions on the Nature of the GC Lobe}
\label{points}
Here, I synthesize my new analysis with previous work, to draw three broad conclusions about the nature of the lobe.  Each of these conclusions has been disputed before, so it is neccessary to state them clearly before moving on with the discussion.  Using these conclusions, I derive the physical conditions of the GC lobe.

\subsection{The GC Lobe is a Single Object}
\label{single}
The discovery of radio and optical recombination lines, radio continuum, and mid-IR emission in the GC lobe, all with a similar morphology, strengthens the argument that it is a single, coherent object.  The 20 cm radio continuum and optical recombination line emission show a clear bridge between the east and west halves of the GC lobe near $b\sim1$\sdeg, which rejects the possibility that the east and west sides are unrelated phenomena.

As described in \S\ \ref{all_gbtcont}, the 6/20 cm spectral index east and west sides have similar, nonthermal values.  More importantly, the index for both edges of the GC lobe show a similar change with Galactic latitude, with significant steepening toward the highest latitudes.  This similarity suggests that they have a similar physical origin.

Finally, \citet{gcl_recomb} note that the GC lobe has radio recombination line emission with an unusually narrow line width.  The narrow line width implies an unusually low electron temperature for the ionized gas that is not commonly seen in typical \hii\ regions \citep{a96}.  The intensity ratios of the lines detected in the east and west of the GC lobe are similar, implying that all the gas has a similar density and unusually low temperature.

\subsection{The GC Lobe is Composed of Layered Shells}
\label{shellsec}
\begin{figure}[tbp]
\includegraphics[width=\textwidth]{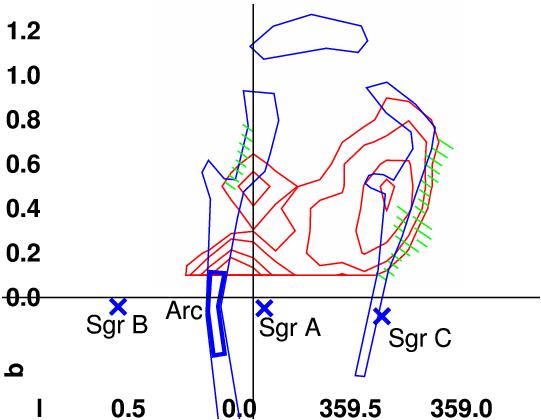}
\caption{Schematic of the multiwavelength structure of the GC lobe and GC region.  The red lines show the contours of radio recombination line emission \citep{gcl_recomb}, the blue schematically shows the radio continuum emission, and the green hatch marks show the mid-IR emission associated with the GC lobe.  The blue crosses show where the brightest \hii\ regions and the thick blue line shows where the brightest emission of the Radio Arc. \label{all_schem}}
\end{figure}

The emission from the three components of the GC lobe have a similar, limb-brightened morphology (see Fig. \ref{all_schem}).  All three components have a similar center longitude, while each component has a different edge-to-edge diameter (see Table \ref{all_shell}).  This gives the appearance nested shells of radio line emission, radio continuum emission, and mid-IR emission.

\citet{b03} model the structure as a shell-like ``telescope dome'' with a radius $r$, height $h$, and thickness $\delta$.  Table \ref{all_shell} shows the best fit shell parameters for five of the observations presented here.\footnote{Parameters from the optical H$\alpha$ are not included, since the emission is extincted and more confused.}  The shell width, $\delta$, is fit by simulating an idealized shell model and convolving it with the telescope beam, $B$;  the thickness is only weakly constrained when it is smaller than the telescope beam.  The height of the GC lobe is measured in the 20 cm radio continuum and optical H$\alpha$ observations to be about 1\ddeg15, or 165 pc at the GC distance.  Under the telescope dome model, the volumes of the radio recombination line, radio continuum, and mid-IR shells are roughly $1.0\times10^6$, $1.9\times10^6$, and $2.8\times10^6$ pc$^{3}$ (assuming equal heights).

To test the shell model, the edge-to-center contrast is measured for each component and compared to that of an idealized model.  I used the simulation to compare the idealized, beam-convolved shell contrast $C_{\rm{ideal}}$ to the observed contrast.  Table \ref{all_shell} shows that three of the observations agree with the idealized shell model, suggesting that the structure seen at those wavelengths may have an intrinsically shell-like structure.  The contrast in the GBT 20 cm continuum is lower than the model predicts, perhaps indicating that the GC lobe is not hollow, but has some 20 cm emission inside.  The contrast of the GBT recombination line structure is larger than predicted, perhaps indicating that the ionized gas is clumpy or irregular.

\subsection{The GC Lobe is in the GC Region}
\label{ingc}
Perhaps the strongest morphological connection between the GC lobe and the GC region is at the Radio Arc.  The present radio continuum survey has confirmed previous work that found contiguous emission from the east of the lobe through the Radio Arc \citep[shown schematically in Fig. \ref{all_schem};][]{y88}.  The Radio Arc (and other nonthermal radio filaments) are known to be within the central few hundred parsecs of the Galaxy from \hi\ absorption measurements \citep{l89,r03}.  This implies that the GC lobe is also in the central few hundred parsecs.

Radio recombination line observations have also found an unusually low electron temperature that implies a high metal abundance for the ionized gas in the GC lobe \citep{gcl_recomb}.  The high metallicity for the GC lobe is consistent with the metallicity expected in the GC region, based on the Galactic abundance gradient \citep{a96}.  The detection of optical H$\alpha$ emission near the north of the GC lobe is consistent with a GC distance to the ionized gas.  Also, the thermal gas pressure is similar to pressures observed in the GC region \citep{gcl_recomb}.

The fact that the ionized component of the GC lobe is inside the other components suggests it is ionized from the inside.  This is consistent with a GC location, since the ionizing flux of massive stars in the central tens of parsecs are sufficient to ionize the gas \citep{gcl_recomb}.

Taken together, these observations show that the GC lobe is located in the central few hundred parsecs of our Galaxy.  Thus, the appearance of the GC lobe spanning the central 100 pc in projection is likely to be true physically.

\subsection{The Physical Conditions of the GC Lobe}
\label{physcond}
Given that the GC lobe is composed of three nested shells and is located in the GC region, I estimate several of its physical parameters.  As a crude minimum energy required to form the GC lobe, I estimate its gravitational potential.  The brightness-weighted mean height of the radio recombination and mid-IR emission of the GC lobe is about 50 pc north of the plane.  The gravitational potential difference between a galactocentric radius of 1--20 pc and 50 pc is $E_{gr} = 5\times10^{51} * (T_e/3960\ \rm{K})^{0.61}$ ergs, for the ionized gas mass \citep[equivalent to a velocity of about 40 \kms;][]{b91}.  The minimum energy estimate is likely to be an underestimate because the total mass is expected to be at least twice that in the ionized gas \citep[$M_{mol}>3\times10^5$ \msol;][]{b03}.  However, the minimum energy may be overestimated by assuming that all the mass originates between 1 and 20 pc.  The true minimum energy is likely to be within an order of magnitude of that calculated above.

The energy required to form the GC lobe can be estimated from the pressure observed throughout it.  The thermal pressure in the ionized gas is measured by the radio recombination line observations to be $P/k=3.8\times10^6 (T_e/3960\ \rm{K})$ K cm$^{-3}$ \citep{gcl_recomb}.  For comparison, X-ray--emitting gas in the plane has pressures of $1-5\times10^6$ K cm$^{-3}$ \citep{k96, m04}, molecular gas velocity dispersion implies virial pressure of $6\times10^6$ K cm$^{-3}$, and the equipartition magnetic field implied by the radio synchrotron emission has a pressure of $1\times10^6$ K cm$^{-3}$.  For an adiabatically expanding shell, the formation energy is $E=\frac{\gamma}{\gamma-1}PV$, where $V$ is the shell volume and $\gamma$ equals $5/3$ if the gas is nonrelativistic and $4/3$ if relativistic \citep{d04}.  The present observations have detected relativistic (synchrotron) and nonrelativistic (recombination line) emission from inside the shell, but have not measured the pressure from that region; it is not clear that one dominates the internal pressure.  Assuming the volume enclosed by the mid-IR shell of the GC lobe, the formation of the GC lobe requires $(4-6)\times10^{52} (T_e/3960\ \rm{K})$ ergs, depending if it is filled with thermal or nonthermal plasma.

This formation energy is consistent with the minimum (gravitational) energy in the GC lobe.  This energy estimate does not account for energy losses \citep{v94} or the kinetic energy associated with the expansion of the GC lobe.  The kinetic energy estimate given in \citet{b03} was about an order of magnitude larger than this formation energy, for an assumed expansion velocity of 100 \kms. \footnote{Note that this previous work calculated a total mass for the GC lobe an order of magnitude larger than the ionized mass found in the present work.  This relies on the analysis of molecular line data that covers Galactic latitudes less than $0\ddeg3$, in which emission from the plane is highly confused with emission that may be from the GC lobe \citep{s95}.}

The upper limit on the line-of-sight expansion of the GC lobe \citep[$\lesssim10$ \kms; ][]{gcl_recomb} gives a dynamical time, $t_{\rm{dyn}}\gtrsim10$ Myr, assuming constant expansion velocity.  If the expansion of the gas has decelerated, the dynamical time will overestimate the actual formation time.  Assuming an initial expansion velocity, $v_{init}$, decelerating to 10 \kms, the formation time is $t_{form} \approx 2 (\Delta r/50$\ pc$)(40$\ km s$^{-1}/v_{init})$ Myr.  The contiguous shape of the GC lobe and its modest height of $\approx165$ pc \citep[only roughly twice the \hi\ scale height in the nuclear disk;][]{r82} show that it has not blown out of the disk.  

Without a measurement on the motion of the GC lobe away from the plane, it is difficult to constrain its fate.  The line-of-sight expansion is certainly far less than the escape velocity \citep[$v_{esc}\approx900$ \kms;][]{b91}.  If the energy source that created the GC lobe continues to operate, then it may expand further or even accelerate, depending on the history of the energy input \citep{v05}.  If the GC lobe is formed by the expansion of hot gas, the ``escape temperature'' is $T_{\rm{esc}}=1.1\times10^5 (v_{\rm{esc}}/100$ \kms$) \rm{K} \approx 9\times10^6\ \rm{K}$  \citep{w95}, which is approximately 1 keV.  However, no hot, X-ray gas has yet been found directly associated the GC lobe.

Finally, it is worth noting a possible constraint on the angular momentum of the ionized gas in the GC lobe.  Radio recombination line velocities show a trend for the ionized gas in the GC lobe to rotate like the disk gas \citep{gcl_recomb}, suggesting that the gas came from the disk \citep[perhaps entrained, as seen in extragalactic outflows;][]{s98,wa02}.  The gas velocity changes across the central 100 pc by 5 \kms, as compared to molecular gas rotation of 200 \kms\ \citep{b87,s04}.  In extragalactic outflows, conservation of angular momentum reduces the velocity gradient from galactic rotation during expansion \citep{s98,s01}.  If this happens in the GC lobe, then the current angular momentum implies that the gas originated at a radius, $r_o \approx 2.5\ \rm{km\ s}^{-1} * 40\ \rm{pc}/100\ \rm{km\ s}^{-1} = 1\ \rm{pc}$.  This value would be an underestimate of the initial radius if the gas velocities were less than $\pm100$\ \kms \citep[e.g., on ``x2'' orbits;][]{bi91} or angular momentum was lost during expansion.

\section{Discussion}
\label{discussion}

\subsection{The Starburst Model for the GC lobe}
\label{starburst}
I now show how well existing models describe the observations of the outflow.  First I consider the canonical starburst outflow model, one of the most widely observed origins of outflows in the local universe \citep{v05}.

The range of estimated energies required to create the GC lobe is consistent with the energy input by supernovae and stellar winds in the GC region.  The gravitational and thermodynamic energy estimates range from an equivalent of $(25-250)/\xi_{0.2}$ of the canonical, $10^{51}$-erg, type-II supernovae, where $\xi_{0.2}$ is the thermalization efficiency scaled to a value of 0.2 \citep{st03,v05}.  The supernova rate in the GC region has been estimated to be roughly $10^{-5}$ yr$^{-1}$ by (1) the need for an energy source for the diffuse, 0.8-keV gas in the GC region and (2) scaling the Galactic supernova rate to the central 20 pc by mass \citep{m04,la02}.  For a formation time of about $10^7$ years, of order 100 supernovae are expected to occur in the central 20 pc.  So the energy required to form the GC lobe is comparable to that expected from GC supernovae during its formation.

Stellar winds are also a significant source of power in the GC region.  \citet{c92} noted that the hot stars in the central parsec of the Galaxy have a wind power of roughly $5\times10^{37}$ ergs s$^{-1}$.  The dozen or so stars considered in that work could power the GC lobe if it operates for $1.6\times10^8/\xi_{0.2}$ yr.  The mass loss by windy stars observed in the central parsecs is now known to be a few times larger and includes massive clusters of stars \citep{g05,f99}.  Thus, the total stellar wind energy input to the GC lobe (and hence, time scale for its formation) is comparable to that of GC supernovae.  Interestingly, since the current stars and supernovae can power the GC lobe, no dramatic increase in star formation rate is needed to explain its formation.  The term ``starburst'' may be an overdramatic description of what formed the lobe.

The observed pressures from the central parsec out to the GC lobe are also consistent with a model powered by stellar winds in the central parsecs \citep{h90,c92}.  The thermal pressure in the GC lobe is consistent with the trend noted in \citet{c92}, which found $P/k\approx3\times10^6$ K cm$^{-3}$ outside of $r\approx4$ pc and rising 2 orders of magnitude in the central parsec.  In this model, the shell of the GC lobe represents the contact discontinuity where the outflow meets the ISM.  The model predicts that the magnetic pressure will begin to dominate the total pressure outside the termination shock, which could explain the sychrotron-emitting shell of the GC lobe.  The estimated equipartition magnetic field pressure is about 1/3 the thermal pressure in the GC lobe, but they may in fact be in equilibrium if the gas has a low filling factor or high proton-to-electron energy density ratio \citep{la05,b05}.

The small gas velocities ($\lesssim10$ \kms) in the GC lobe do not exclude any outflow models.  While typical edge-on extragalactic outflows have emission line-of-sight expansion of more than 200 \kms\ \citep{h90}, the GC lobe is in a much different environment.  \citet{m05} observe starburst outflow velocities that scale roughly as $SFR^{0.35}$, which implies that a putative GC outflow would be several times smaller than canonical outflows \citep[e.g., $(5/0.02)^{0.35}\approx7$ times smaller than M82;][]{h90,f04}.  Furthermore, the expansion velocity of the GC lobe might be expected to be relatively small, since extragalactic outflows tend to be collimated near the galactic plane \citep{c85,su94}.  In one case where the M82 outflow was observed on 200 pc size scales, the outflow velocity was about $\sim50$ \kms\ \citep{s98}.  Alternatively, the low velocity of the gas is also consistent with the ``neutral outflow'' model associated with individual super star clusters \citep{t06};  this is discussed futher in \S \ref{layered}

Because of its small size, the GC lobe is difficult to compare to other outflows.  A good extragalactic analogue of the GC lobe might be the outflow in the nearest starburst galaxy, the dwarf irregular IC 10 \citep{t05}.  That outflow is composed of a nonthermal radio continuum lobe with a diameter of about 200 pc that is filled with ionized gas \citep{y93,t05}.  The IC 10 superbubble is also similar to the GC lobe in that it has no clear sign of expansion; it has a kinetic (turbulent) energy of $5\times10^{52}$ ergs and age of roughly $7\times10^6$ years.

\subsection{Other Models}
\label{other}
Active galactic nuclei (AGN) are also often observed to power nuclear mass outflows \citep{v05}.  The black hole at the center of the Milky Way, Sgr A*, has a mass comparable to that seen in typical AGN \citep[e.g.,][]{fe01}, suggesting that Sgr A* can emit intense radiation and launch powerful jets.  However, jet-powered outflows tend to be narrower near the disk than beyond, where interaction with ambient material reduces the outflow momentum and the ambient gas pressure is lower \citep{v05}.  The width of the GC lobe is similar in the plane as away from the plane and is brighter at its eastern and western edges than at its apex, which is inconsistent with the canonical jet-powered outflows \citep[e.g., ][]{p84}.

Another possibility to power a GC mass outflow is via the escape of its pervasive hot plasma \citep{k96,m04}.  X-ray observations have found a very hot, 8 keV gas throughout the GC region.  Its origin is not known, but it is too energetic to remain bound to the Galaxy \citep{m04,be05}.  It is possible that as this gas escapes from the Galaxy, it creates the structure seen as the GC lobe.  Unfortunately, testing the 8 keV gas model is difficult, since the origin of the gas is not well known.  One possible source for heating the gas is viscous heating by molecular clouds, which could power the GC lobe if it operated for $3\times10^7$ yr \citep{be05,be06};  this time is similar to the (uncertain) age of the GC lobe.  In most respects, the escape of 8 keV gas would have similar signatures as the starburst model, since they both involve the buoyant escape of overpressurized gas.  One possible way to distinguish between the two models may be to search for 8 keV gas inside the GC lobe, but distinct from the Galactic plane.

Finally, we consider that the GC lobe is powered by a magnetodynamic effect called the ``magnetic twist'' \citep{u85,su87}.  Simulations have shown that the interaction of orbiting gas clouds in a predominately poloidal (vertical) magnetic field can create a pinched magnetic field configuration.  This configuration can push ionized gas away from the Galactic disk by the ${\bf J}\times {\bf B}$ force, possibly creating an outflow \citep{u85,su87}.  Observations have found that the magnetic field in the GC region is perturbed by dense gas, consistent with the magnetic twist model \citep{y88,c03}.  However, the magnetic twist mechanism was developed to duplicate the morphology of the GC lobe, so a quantitative comparison to observations is difficult.  Also, it isn't clear how the model would produce the layered structure of the GC lobe.  Thus, more work is needed before applying the magnetic twist model to the formation of the GC lobe.

\subsection{Remaining Questions}
\subsubsection{The Layered Outflow}
\label{layered}
While the GC lobe clearly has layers of different types of emission, the reason for this layered structure is not clear.  Central to understanding this structure is knowing whether the radio and optical recombination line emission is shock- or photo-ionized.

As discussed in \citet{gcl_recomb}, the number of ionizing photos from hot stars in the GC region can explain the radio recombination line emission.  If the GC lobe is photoionized, the offset in the ionized and dusty layers implies that the ionization can't reach the dust layer because it is ionization bounded \citep{r09}.  In this case the GC lobe may be thought of as a massive neutral outflow, such as observed around super star clusters \citep{g00,t06}.  As these outflows are much less energetic than supersonic starburst winds, they are not shock ionized or heated enough to produce soft X-rays.

A supersonic outflow, typical of extragalactic starbursts, may also explain the layered structure of the GC lobe.  Simulations of supersonic winds have shown how they tear ambient clouds into shock-ionized filaments that emit H$\alpha$ and soft X-rays \citep{c09,v94,s98}.  If the GC lobe is shock ionized, the ionized layer may trace the portion of the outflow that has had time to ionize, which occurs over several times $10^5$ years \citep{c09}.  Meanwhile, at the contact discontinuity, the gas is piled up, but does not yet emit radio or optical recombination lines.  

These two models for the GC lobe can be distinguished by their soft X-ray emission or line ratios of ionized species.  If the outflow is supersonic, the gas is shock ionized and soft X-ray emission is expected to be coincident with the H$\alpha$ emission.  If no soft X-ray emission is found coincident with the H$\alpha$ emission, the gas must be photoionized.  Alternatively, the ratio of certain forbidden and recombination lines can be modeled to measure the amount of shock heating in the ionized gas \citep{v05}.

\subsubsection{AFGL 5376}
AFGL 5376 is an extended infrared source discovered by \emph{IRAS} near the GC that seems to be a shocked molecular cloud \citep{u90,u94}.  \citet{u94} show that radio continuum emission of the GC lobe is coincident with a shock in AFGL 5376, suggesting that they are related.

While the morphology suggests that these sources are interacting, the direction of causality is unclear.  \citet{u94} note that the shock in the molecular gas in AFGL 5376 could ionize it and generate the radio continuum of the western part of the GC lobe.  While the present data confirm that there is thermal radio continuum emission near AFGL 5376, it is also part of a structure that is predominately nonthermal.  The causal relation between AFGL 5376 and the GC lobe, is more clear when considering the radio recombination line emission found throughout the region.  Since there is no way for AFGL 5376 to create the ionized gas that fills the GC lobe, the GC lobe must exist independently of AFGL 5376.

One possibility is that AFGL 5376 is the result of a collision between a molecular cloud with the GC lobe.  The molecular gas in AFGL 5376 is peculiar in that it is massive ($10^6$ \msol) and moves at more than 122 km s$^{-1}$ counter to Galactic rotation \citep{u94};  this makes it likely to have an energetic collision with ambient, co-rotating gas.  If this is an overrun cloud, high resolution ($<$1 pc) observations of an ionized gas tracer may find filaments of emission, as seen in simulations \citep{c09} and other starburst outflows \citep{s98}.

\subsubsection{Double Helix Nebula}
Recent mid-IR observations with \spitzer\ have found a pair of twisted filaments about 25 pc in length called the ``double helix nebula'' \citep[DHN;][]{m06}.  The twisting of the DHN strongly suggests that its structure is dominated by magnetic forces.  The DHN is also clearly associated with the GC lobe because it makes up the brightest mid-IR emission in its eastern half.  The model proposed for the DHN is that of a Alfv\'en wave propagating from the molecular gas in the central parsec of the Galaxy and is unrelated to the GC lobe \citep{m06}.

The nature of the DHN is still unclear and it might not conflict with the outflow model for the GC lobe.  The model of the DHN as an Alfv\'en wave requires propagation from the central parsec, but there is little morphological evidence for such a propagation.  \citet{m06} identify two features in the 8 $\mu$m \msx\ map of the GC region that could indicate a path, but at least one of the features is also seen in optical H$\alpha$ emission and is thus likely to be in the foreground \citep{s06}.  Indeed, the idea that the DHN is created by magnetic forces is consistent with its location in the edge of the GC lobe, where, as discussed in \S \ref{starburst}, we expect a stronger magnetic field.  An alternate possibility is that the wavy, filamentary structure of the DHN is similar to the structures seen in optical emission lines at the shock fronts of the SNRs \citep[e.g., the Cygnus Loop;][]{bl99}.  These structures arise naturally as the shock front propagates into an inhomogeneous medium and may shape the shell of the GC lobe.

Interestingly, the DHN is adjacent to a wavy, 6 cm continuum feature called G0.03+0.66 \citep{gcl_vla}.  G0.03+0.66 has a flat, thermal spectral index.  The nested, thermal radio emission with mid-IR emission is consistent of the general structure of the GC lobe described here, which suggests that the DHN may simply be fine-scale structure in the GC lobe.  If so, high-resolution observations of dust emission will find analogues of the DHN.

\subsubsection{Asymmetries}
\label{asymmetries}
The center of the GC lobe is offset to the west of Sgr A* by $0\ddeg3$--$0\ddeg4\approx40$--$55$ pc.  The starburst outflow model can explain such an offset, since the star formation need not be at the dynamical center \citep{v05}.  An offset like this may also explain the lack of an obvious southern counterpart to the GC lobe, as has been observed in other galaxies \citep{s98}.

The eastern offset of the GC lobe from Sgr A* may also help explain its internal asymmetries.  Assuming that the center of the GC lobe is the center of the outflow, one can imagine that the eastern and western sides of the outflow have encountered drastically different conditions during their expansion.  The large polarization fraction and connection with the unusual Radio Arc in the east of the GC lobe may be the result of the expansion of the outflow through the relatively dense gas and possibly stronger magnetic field in the central few parsecs.

Curiously, the spatial distribution of nonthermal radio filaments has a similar offset as the GC lobe \citep{y04}.  Studies of the distribution of rotation measure values observed toward radio continuum emission in the GC region may also be be offset \citep{n03}.  One possibility is that the outflow has temporarily perturbed the GC magnetosphere from its intrinsic configuration \citep{f09}.  This hypothesis will be discussed in detail elsewehere \citep{gcl_vlapoln}.

\section{Conclusions}
\label{conclusions}
I have conducted a multiwavelength observing program to study the GC lobe and determine if it is a signature of an outflow from the GC region.  The new observations show that the GC lobe has a layered structure with concentric shells of emission of mid-IR, radio continuum, and optical and radio recombination lines.  The morphological, spectral, and physical properties of the GC lobe show that it is a coherent object in the GC region.  I calculate the energy and time required to create the GC lobe and compare its characteristics to models of nuclear outflows.  The canonical starburst model best explains the properties of the GC lobe, although the outflow may not be supersonic as in typial extragalactic starburst outflows.  The energy output by current star formation and supernovae can power the GC lobe.

If the GC lobe is indeed a nuclear outflow powered by stellar winds and supernovae, it is the closest example of such a phenomenon.  This opens the possibility of studying an outflow at extremely high physical resolution, potentially expanding our understanding of how they work.  New observations of X-rays or forbidden-line transitions should clarify whether the outflow is supersonic or may be described by a less energetic outflow, such as expected from super star clusters.  The relatively small size of the GC lobe also allows us to explore a type of outflow that is difficult to see in other galaxies.  The fact that the GC lobe would be difficult to detect in even the nearest spiral galaxies suggests that this type of outflow could be common.

The mechanism for periodic infall of mass to the central parsecs \citep{st04} suggests that the GC region may undergo cycles of activity \citep{a00,y04}.  It may be that the GC lobe is just the most recent example of a long series of Milky Way nuclear starburst outflows.

\acknowledgements{I am grateful to Martin Cohen, Farhad Yusef-Zadeh, and Jessica Law for valuable discussions.  I thank the referee for constructive comments that improved the paper.  The National Radio Astronomy Observatory is a facility of the National Science Foundation operated under cooperative agreement by Associated Universities, Inc.  I acknowledge the Southern H-Alpha Sky Survey Atlas (SHASSA), which is supported by the National Science Foundation, and the Air Force Research Laboratory for the use of the MSX data.  This research has made use of NASA's Astrophysics Data System Bibliographic Services.}

{\it Facilities:} \facility{GBT (), Spitzer (IRAC), }

\clearpage

\begin{deluxetable}{lcccccl}
\tablecaption{Observed and Calculated Parameters for GC lobe, Assuming Shell Morphology \label{all_shell}}
\tablewidth{0pt}
\tablehead{
\colhead{Observation} & \colhead{$r$} & \colhead{$\delta$} & \colhead{$B$} & \colhead{$C_{\rm{ideal}}$} & \colhead{$C_{obs}$} & \colhead{Notes} \\
\colhead{} & \colhead{(pc)} & \colhead{(pc)} & \colhead{(pc)} & \colhead{} & \colhead{} & \colhead{} \\
}
\startdata
HCRO Recombination line & 40 & 15 & 23   & 1.5 & 1.5--3 & at $b=0\ddeg5$ \\
GBT Recombination line  & 40 & 15 & 5.8  & 2.1 & 3--7 & at $b=0\ddeg45$ \\
GBT 6 cm Continuum      & 55 & 15 & 5.8  & 2.4 & 2.2--2.6 & at $b=0\ddeg45$ \\
GBT 20 cm Continuum     & 55 & 15 & 21   & 1.8 & 1.2--1.4 & at $b=0\ddeg45$ \\
\emph{Spitzer} 8$\mu$m  & 65 &  5 & 0.08 & 5.2 & $>3$\tablenotemark{a} & at $b=0\ddeg56$ \\
\enddata
\tablenotetext{a}{The emission from the center of the GC lobe seems to be mostly foreground emission.  Thus, the observed contrast is a lower limit to the contrast associated with the shell itself.}
\end{deluxetable}

\clearpage

\begin{thebibliography}{}
\bibitem[Afflerbach et al.(1996)]{a96} Afflerbach, A. Churchwell, E., Acord, J. M., Hofner, P., Kurtz, S., \& Dupree, C. G. 1996, ApJS, 106, 423

\bibitem[Allamandola et al.(1989)]{a89} Allmandola, L. J., Tielens, A. G. G. M., \& Barker, J. R. 1989, ApJS, 71, 733

\bibitem[Almy et al.(2000)]{a00} Almy, R. C., McCammon, D., Digel, S. W., Bronfman, L., \& May, J. 2000, ApJ, 545, 290

\bibitem[Arendt et al.(2008)]{a08} Arendt, R. G. 2008, ApJ, 682, 384



\bibitem[Bally et al.(1987)]{b87} Bally, J., Stark, A. A., Wilson, R. W., \& Henkel, C. 1987, ApJS, 65, 13

\bibitem[Beck \& Krause(2005)]{b05} Beck, R. \& Krause, M. 2005, AN, 326, 414

\bibitem[Belmont et al.(2005)]{be05} Belmont, R., Tagger, M., Muno, M., Morris, M., \& Cowley, S. 2005, ApJ, 631, L53

\bibitem[Belmont \& Tagger(2006)]{be06} Belmont, R. \& Tagger, M. 2006, A\&A, 452, 15


\bibitem[Binney et al.(1991)]{bi91} Binney, J., Gerhard, O. E., Stark, A. A., Bally, J., \& Uchida, K. I. 1991, MNRAS, 252, 210

\bibitem[Blair et al.(1999)]{bl99} Blair, W. P., Sankrit, R., Raymond, J. C., \& Long, K. S. 1999, AJ, 118, 942

\bibitem[Bland-Hawthorn \& Cohen(2003)]{b03} Bland-Hawthorn, J \& Cohen, M. 2003, ApJ, 582, 246




\bibitem[Breitschwerdt et al.(1991)]{b91} Breitschwerdt, D., Voelk, H. J., \& McKenzie, J. F. 1991, A\&A, 245, 79

\bibitem[Brogan et al.(2003)]{b03b} Brogan, C. L., Nord, M., Kassim, N., Lazio, J., \& Anantharamaiah, K. 2003, ANS, 324, 17

\bibitem[Cardelli, Clayton, \& Mathis(1989)]{c89} Cardelli, J. A., Clayton, G. C., \& Mathis, J. S. 1989, ApJ, 345, 245


\bibitem[Chevalier(1992)]{c92} Chevalier, R. A. 1992, ApJ, 397, L39

\bibitem[Chevalier \& Clegg(1985)]{c85} Chevalier, R. A. \& Clegg, A. W. 1985, Nature, 317, 44


\bibitem[Chuss et al.(2003)]{c03} Chuss, D. T., Davidson, J. A., Dotson, J. L., Dowell, C. D., Hildebrand, R. H., Novak, G., \& Vaillancourt, J. E. 2003, ApJ, 599, 1116

\bibitem[Cooper et al.(2009)]{c09} Cooper, J. L., Bicknell, G. V., Sutherland, R. S., Bland-Hawthorn, J. 2009, ApJ, 703, 330


\bibitem[Dunn \& Fabian(2004)]{d04} Dunn, R. J. H., \& Fabian, A. C. 2004, MNRAS, 355, 862

\bibitem[Dutra et al.(2003)]{d03} Dutra, C. M., Santiago, B. X., Bica, E. L. D., \& Barbuy, B. 2003, MNRAS, 338, 253





\bibitem[Ferrarese et al.(2001)]{fe01} Ferrarese, L., Pogge, R. W., Peterson, B., M., Merritt, D., Wandel, A., \& Joseph, C. L. 2001, ApJ, 555L, 79

\bibitem[Ferri\'ere(2009)]{f09} Ferri\'ere, K. 2009, astro-ph/0908.2037

\bibitem[Figer et al.(2004)]{f04} Figer, D. F., Rich, R. M., Kim, S. S., Morris, M., \& Serabyn, E. 2004, ApJ, 601, 319

\bibitem[Figer et al.(1999)]{f99} Figer, D. F., McLean, I. S. \& Morris, M. 1999, ApJ, 514, 202


\bibitem[Gaustad et al.(2001)]{g01} Gaustad, J. E., McCullough, P. R., Rosing, W., \& Van Buren, D. 2001, PASP, 113, 1326

\bibitem[Ghez et al.(2005)]{g05} Ghez, A., et al. 2005, ApJ, 620, 744

\bibitem[Gonz\'alez Delgado \& P\'erez(2000)]{g00} Gonz\'alez Delgado, R. M. \& P\'erez, E. 2000, MNRAS, 317, 64


\bibitem[Gossling, Bandyopadhyay, \& Blundell(2009)]{g09} Gosling, A. J., Bandyopadhyay, R. M., \& Blundell, K. M. 2009, MNRAS, 394, 2247


\bibitem[Haynes et al.(1992)]{h92} Haynes, R. F., Stewart, R. T., Gray, A. D., Reich, W., Reich ,P., \& Mebold, U. 1992, A\&A, 264, 500

\bibitem[Heckman et al.(1990)]{h90} Heckman, T. M., Armus, L., \& Miley, G. K. 1990, ApJS, 74, 833

\bibitem[Hopkins et al.(2005)]{h05} Hopkins, P. F., Hernquist, L., Cox, T. J., Di Matteo, T., Martini, P., Robertson, B., \& Springel, V. 2005, ApJ, 630, 705


\bibitem[Koyama et al.(1996)]{k96} Koyama, K., Maeda, Y., Sonobe, T., Takeshima, T., Tanaka, Y., \& Yamauchi, S. 1996, PASJ, 48, 249

\bibitem[Lang et al.(2001)]{l01} Lang, C. C., Goss, W. M., \& Morris, M. 2001, AJ, 121, 2681

\bibitem[LaRosa et al.(2005)]{la05} LaRosa, T. N., Brogan, C. L., Shore, S. N., Lazio, T. J., Kassim, N. E., \& Nord, M. E.  2005, ApJ, 626, L23


\bibitem[Launhardt et al.(2002)]{la02} Launhardt, R., Zylka, R., \& Mezger, P. G. 2002, A\&A, 384, 112

\bibitem[Lasenby et al.(1989)]{l89} Lasenby, J., Lasenby, A. N., \& Yusef-Zadeh, F. 1989, ApJ, 343, 177

\bibitem[Law et al.(2008a)]{gcl_vla} Law, C. J., Yusef-Zadeh, F., \& Cotton, W. D. 2008a, ApJS, 177, 515

\bibitem[Law et al.(2008b)]{gcsurvey_gbt} Law, C. J., Yusef-Zadeh, F., Cotton, W. D., \& Maddalena, R. J. 2008b, ApJS, 177, 255

\bibitem[Law et al.(2009)]{gcl_recomb} Law, C. J., Backer, D., Yusef-Zadeh, F., \& Maddalena, R. 2009, ApJ, 695, 1070

\bibitem[Law et al.(2010)]{gcl_vlapoln} Law, C. J., et al. 2010, in preparation



\bibitem[Madsen \& Reynolds(2005)]{mad05} Madsen, G. J. \& Reynolds, R. J. 2005, ApJ, 63, 925

\bibitem[Markoff(2005)]{mar05} Markoff, S. 2005, ApJ, 618L, 103

\bibitem[Martin(2005)]{m05} Martin, C. L. 2005, 621, 227



\bibitem[Melia \& Falcke(2001)]{m01} Melia, F. \& Falcke, H. 2001, ARA\&A, 39, 309

\bibitem[Morris \& Serabyn(1996)]{m96} Morris, M. \& Serabyn, E. 1996, ARA\&A, 34, 645

\bibitem[Morris et al.(2006)]{m06} Morris, M., Uchida, K., \& Do, T. 2006, Nature, 440, 308

\bibitem[Muno et al.(2004)]{m04} Muno, M. P. et al. 2004, ApJ, 613, 326

\bibitem[Novak et al.(2003)]{n03} Novak, G., et al. 2003, ApJ, 583L, 83

\bibitem[Peeters et al.(2004)]{p04} Peeters, E., Spoon, H. W. W., \& Tielens, A. G. G. M. 2004, ApJ, 613, 986

\bibitem[Perley et al.(1984)]{p84} Perley, R. A., Dreher, J. W., \& Cowan, J. J. 1984, ApJ, 285L, 35

\bibitem[Pohl et al.(1992)]{p92} Pohl, M., Reich, W., \& Schlickeiser, R. 1992, A\&A, 262, 441

\bibitem[Price et al.(2001)]{p01} Price, S. D., Egan, M. P., Carey, S. J., Mizuno, D. R., \& Kuchar, T. A. 2001, AJ, 121, 2819

\bibitem[Ramirez et al.(2008)]{r08} Ramirez, S. 2008, ApJS, 175, 147

\bibitem[Reich et al.(1987)]{r87} Reich, W., Sofue, Y., \& Fuerst, E. 1987, PASJ, 39, 573

\bibitem[Reid(1993)]{r93} Reid, M. J. 1993, ARA\&A, 31, 345


\bibitem[Rohlfs \& Braunsfurth(1982)]{r82} Rohlfs, K. \& Braunsfurth, E. 1982, A\&A, 113, 237

\bibitem[Rodr\'iguez et al.(2009)]{r09} Rodríguez, L. F., G\'omez, Y., \& Guzm\'an, L. 2009, RMxAA, 45, 85

\bibitem[Roy(2003)]{r03} Roy, S. 2003, A\&A, 403, 917


\bibitem[Sawada et al.(2004)]{s04} Sawada, T., Hasegawa, T., Handa, T., \& Cohen, R. J. 2004, MNRAS, 349, 1167


\bibitem[Seaquist \& Clark(2001)]{s01} Seaquist, E. R. \& Clark, J. 2001, ApJ, 552, 133

\bibitem[Shibata \& Uchida(1987)]{su87} Shibata, K. \& Uchida, Y. 1987, PASJ, 39, 559

\bibitem[Shopbell \& Bland-Hawthorn(1998)]{s98} Shopbell, P. L. \& Bland-Hawthorn, J. 1998, ApJ, 493, 129


\bibitem[Sofue(1985)]{s85} Sofue, Y. 1985, PASJ, 37, 697

\bibitem[Sofue(1995)]{s95} Sofue, Y. 1995, PASJ, 47, 527



\bibitem[Sofue \& Handa(1984)]{s84} Sofue, Y. \& Handa, T. 1984, Nature, 310, 568



\bibitem[Stark et al.(2004)]{st04} Stark, A. A., Martin, C. L., Walsh, W. M., Xiao, K., Lane, A. P., Walker, C. K. 2004, ApJ, 614L, 41


\bibitem[Stevens \& Hartwell(2003)]{st03} Stevens, I. R. \& Hartwell, J. M. 2003, MNRAS, 339, 280

\bibitem[Stolovy et al.(2006)]{s06} Stolovy, S., et al. 2006, JPhCS, 54, 176

\bibitem[Suchkov et al.(1994)]{su94} Suchkov, A. A., Balsara, D. S., Heckman, T. M., \& Leitherner, C. 1994, ApJ, 430, 511



\bibitem[Tenorio-Tagle et al.(2006)]{t06} Tenorio-Tagle, G., Mu\~noz-Tu\~n\'on, C., Pérez, E., Silich, S., \& Telles, E. 2006, ApJ, 643, 186

\bibitem[Thurow \& Wilcots(2005)]{t05} Thurow, J. C. \& Wilcots, E. M. 2005, AJ, 129, 745

\bibitem[Tsuboi et al.(1986)]{t86} Tsuboi, M., et al. 1986, AJ, 92, 818

\bibitem[Uchida et al.(1990)]{u90} Uchida, K., Morris, M., \& Serabyn, E. 1990, ApJ, 351, 443

\bibitem[Uchida et al.(1994)]{u94} Uchida, K. I., et al.\ 1994, ApJ, 421, 505

\bibitem[Uchida et al.(1985)]{u85} Uchida, Y., Shibata, K., \& Sofue, Y. 1985, Nature, 317, 699

\bibitem[Veilleux et al.(2005)]{v05} Veilleux, S., Cecil, G., \& Bland-Hawthorn, J. 2005, ARA\&A, 43, 769

\bibitem[Veilleux et al.(1994)]{v94} Veilleux, S., Cecil, G., Bland-Hawthorn, J., Tully, R. B., Filippenko, A. V., \& Sargent, W. L. W. 1994, ApJ, 433, 48

\bibitem[Walter et al.(2002)]{wa02} Walter, F., Weiss, A., \& Scoville, N. 2002, ApJ, 580, L21

\bibitem[Wang(1995)]{w95} Wang, B. 1995, ApJ, 444, 590

\bibitem[Yao \& Wang(2007)]{y07} Yao, Y. \& Wang, Q. D. 2007, ApJ, 666, 242

\bibitem[Yang \& Skillman(1993)]{y93} Yang, H. \& Skillman, E. D. 1993, AJ, 106, 1448


\bibitem[Yusef-Zadeh \& Morris(1988)]{y88} Yusef-Zadeh, F. \& Morris, M. 1988, ApJ, 329, 729

\bibitem[Yusef-Zadeh et al.(1984)]{y84} Yusef-Zadeh, F., Morris, M., \& Chance, D. 1984, Nature, 310, 557

\bibitem[Yusef-Zadeh et al.(2004)]{y04} Yusef-Zadeh, F., Hewitt, J., \& Cotton, W.  2004, ApJS, 155, 421

\bibitem[Yusef-Zadeh et al.(2006)]{y06} Yusef-Zadeh, F., et al. 2006, ApJ, 644, 198


\end{thebibliography}
\end{document}